\documentclass[%
 reprint,
 amsmath,amssymb,
 aps,
]{revtex4-2}

\usepackage{graphicx}
\usepackage{dcolumn}
\usepackage{bm}
\usepackage[dvipsnames,table]{xcolor} %
\usepackage{physics}
\usepackage{braket}
\usepackage[breaklinks, %
colorlinks, linkcolor=Blue, citecolor=Blue, urlcolor=Blue]{hyperref} %
\usepackage{tabularx} %
\usepackage{booktabs} %
\usepackage{float} %
\newcommand{\bea}{\begin{eqnarray}}
\newcommand{\eea}{\end{eqnarray}}
\begin{document}

\preprint{APS/123-QED}

\title{Exactly solvable Hamiltonian fragments obtained from a direct sum of Lie algebras}

\author{Smik Patel}
\affiliation{Chemical Physics Theory Group, Department of Chemistry, University of Toronto, Toronto, Ontario M5S 3H6, Canada}
\affiliation{Department of Physical and Environmental Sciences, University of Toronto Scarborough, Toronto, Ontario M1C 1A4, Canada}
\author{Artur F. Izmaylov}
\affiliation{Chemical Physics Theory Group, Department of Chemistry, University of Toronto, Toronto, Ontario M5S 3H6, Canada}
\affiliation{Department of Physical and Environmental Sciences, University of Toronto Scarborough, Toronto, Ontario M1C 1A4, Canada}

\date{\today}

\begin{abstract}

Exactly solvable Hamiltonians are useful in the study of quantum many-body systems using quantum computers. In the variational quantum eigensolver, a decomposition of the target Hamiltonian into exactly solvable fragments can be used for evaluation of the energies via repeated quantum measurements. In this work, we apply more general classes of exactly solvable qubit Hamiltonians than previously considered to address the Hamiltonian measurement problem. The most general exactly solvable Hamiltonians are defined by the condition that, within each simultaneous eigenspace of a set of Pauli symmetries, the Hamiltonian acts effectively as an element of a direct sum of $so(N)$ Lie algebras, and can therefore be measured using a combination of unitaries in the associated Lie group, Clifford unitaries, and mid-circuit measurements. Application of such Hamiltonians to decomposing molecular electronic Hamiltonians via graph partitioning techniques shows a reduction in the total number of measurements required to estimate the expectation value compared with previously used exactly solvable qubit Hamiltonians.

\end{abstract}

\maketitle


\section{\label{intro_sec}Introduction}

Exactly solvable Hamiltonians are of fundamental importance in the study of quantum many-body systems on quantum computers. In quantum simulation algorithms, obtaining a decomposition of the target Hamiltonian $\hat{H} = \sum_\alpha \hat{H}_\alpha$ into exactly solvable fragments is a crucial part of the Hamiltonian encoding for near-term \cite{huggins2019efficient, yenCartanSubalgebraApproach2021, Choi2023fluidfermionic, zhaoMeasurementReductionVariational2020, izmaylovUnitaryPartitioningApproach2020, verteletskyiMeasurementOptimizationVariational2020, yenMeasuringAllCompatible2020, patelExtensionExactlysolvableHamiltonians2023, JenaFC, crawfordEfficientQuantumMeasurement2021} and fault-tolerant
\cite{lloydUniversalQuantumSimulators1996, childsHamiltonianSimulationUsing2012, loaizaReducingMolecularElectronic2023a,martinez-martinezAssessmentVariousHamiltonian2023}
algorithms.  For the variational quantum eigensolver (VQE), \cite{peruzzoVariationalEigenvalueSolver2014} one approach to compute $\langle \hat{H} \rangle$ via quantum measurements is to sum estimates of $\langle \hat{H}_\alpha \rangle$ which can be measured directly due to their exact solvability. This motivates the search for new, more general classes of exactly solvable Hamiltonians that can be used for Hamiltonian partitionings. In this work, we focus on decompositions of the electronic Hamiltonian in the second-quantized form
\begin{equation}
    \hat{H}_e = \sum_{pq=1}^{N} h_{pq} \hat{a}_p^\dagger \hat{a}_q + \sum_{pqrs=1}^{N} g_{pqrs} \hat{a}_p^\dagger \hat{a}_q \hat{a}_r^\dagger \hat{a}_s,
\end{equation}
where $h_{pq}$ and $g_{pqrs}$ are one- and two-electron integrals, and $p,q,r,s$ enumerate the $N$ spin-orbitals. \cite{helgakerMolecularElectronicStructureTheory2014}  $\hat{H}_e$ can also be expressed as a linear combination of $N$-qubit Pauli operators via a fermion-to-qubit mapping like the Jordan-Wigner \cite{jordanUeberPaulischeAequivalenzverbot1928} or Bravyi-Kitaev transformations. \cite{bravyiFermionicQuantumComputation2002, seeleyBravyiKitaevTransformationQuantum2012} 

Previous works on the Hamiltonian partitioning problem when applied to the electronic Hamiltonian have exploited the algebraic structure of quantum many-body operators to obtain simple exactly solvable models. \cite{huggins2019efficient, zhaoMeasurementReductionVariational2020, izmaylovUnitaryPartitioningApproach2020, yenCartanSubalgebraApproach2021, Choi2023fluidfermionic, patelExtensionExactlysolvableHamiltonians2023, yenMeasuringAllCompatible2020, verteletskyiMeasurementOptimizationVariational2020} These approaches rely on the maximal torus theorem, which implies that Hamiltonians that are elements of a compact Lie algebra can be rotated to a linear combination of mutually commuting operators by a unitary in the corresponding Lie group. \cite{hallLieGroupsLie2015} In the fermionic picture, number-conserving free-fermionic Hamiltonians, which are linear combinations of $u(N)$ algebra generators $\{i(\hat{a}_p^\dagger \hat{a}_q - \hat{a}_q^\dagger \hat{a}_p)\} \cup \{\hat{a}_p^\dagger \hat{a}_q + \hat{a}_q^\dagger \hat{a}_p\}$ on $N$ spin-orbitals, are exactly solvable. From this, one can form exactly solvable fragments of $\hat{H}_e$ which are quadratic polynomials of $u(N)$ elements $\{\hat{U}^\dagger \hat{n}_p\hat{U}\}$, where $\hat{U}$ is an orbital transformation from the associated Lie group, and $\hat{n}_p$ is the number operator for the $p$th spin-orbital. \cite{huggins2019efficient, yenCartanSubalgebraApproach2021, Izmaylov_2021DefMeanField, Choi2023fluidfermionic} In the qubit picture, linear combinations of mutually anti-commuting Pauli operators are exactly solvable, from which the unitary partitioning method was developed. \cite{izmaylovUnitaryPartitioningApproach2020, zhaoMeasurementReductionVariational2020} The exact solvability of such anti-commuting (AC) Hamiltonians follows from the fact that, associated to a set $\{\hat{A}_i\}_{i=1}^{L}$ of mutually anti-commuting Pauli operators, is an $so(L+1)$ Lie algebra generated by $\{\hat{A}_i\}_{i=1}^{L} \cup \{i \hat{A}_i \hat{A}_j\}_{i>j}^{L}$. This motivates the question of whether one can use Hamiltonians constructed from the full set of $so(L+1)$ Lie algebra generators to solve the Hamiltonian partitioning problem. Incorporation of the full set of algebra generators allows for more flexible Hamiltonian fragments, which can provide a reduction in the VQE measurement cost in the electronic structure problem. 

To obtain the fragments in the unitary partitioning approach, graph partitioning algorithms are used. Associated to any qubit Hamiltonian is the anti-compatibility graph, in which the vertices correspond to Pauli terms, and two vertices share an edge if and only if their corresponding Pauli terms anti-commute. The anti-compatibility graph of an AC Hamiltonian is a complete graph, in which every pair of vertices shares an edge. Therefore, the task of finding AC fragments is equivalent to obtaining a partitioning of the electronic Hamiltonian's anti-compatibility graph into complete subgraphs. Graph partitioning algorithms have also been used to obtain a decomposition of the electronic Hamiltonian into fragments whose Pauli terms mutually commute, which are called fully-commuting (FC) fragments \cite{yenMeasuringAllCompatible2020}, and, more recently, fragments whose Pauli terms generate a non-contextual (NC) set. \cite{kirbyContextualityTestNonclassicality2019, patelExtensionExactlysolvableHamiltonians2023}

In this work, we develop the algorithms necessary to partition electronic Hamiltonians into fragments constructed from the full set of $so(L+1)$ algebra generators on a classical computer, and to measure such fragments on a quantum computer. We also consider a simple extension where the Hamiltonian fragments are constructed from Pauli terms that generate a direct sum algebra $\oplus_{\lambda=1}^{\Lambda} so(L_\lambda + 1)$, which can be solved by a product of unitaries from the Lie groups associated to each $so(L_\lambda + 1)$ subalgebra. Our procedures for obtaining the Hamiltonian fragments and deriving the associated quantum measurement circuits relies on the graph theoretical developments of Ref. \cite{chapmanCharacterizationSolvableSpin2020}. Lastly, we benchmark the newly developed fragment types in terms of measurement cost, and compare to the previously used AC,  FC, and NC criterions. 

The rest of this paper is organized as follows. Section \ref{theory_sec} introduces the framework for constructing exactly solvable Hamiltonians using compact Lie algebras and their symmetries, applies this framework to construct exactly-solvable qubit Hamiltonians, and discusses techniques for performing the Hamiltonian partitioning and measurement of the resulting fragments on a quantum computer. Section \ref{results_sec} presents numerical results for decompositions of electronic Hamiltonians into all exactly solvable Hamiltonian types considered in this work. Section \ref{conclusion_sec} concludes and presents outlook.

\section{\label{theory_sec}Theory}

\subsection{\label{liealgebra_subsec} Lie algebraic framework}

Any fermionic or qubit Hamiltonian can be expressed as a polynomial function of a set of elementary Hermitian operators $\{\hat{X}_a\}_{a=1}^{d}$, which generate a $d$ dimensional compact Lie algebra:
\begin{equation}
    \hat{H} = \sum_{a=1}^{d} c_a \hat{X}_a + \sum_{a\geq b}^{d}c_{ab} \hat{X}_a \hat{X}_b + \cdots .
\end{equation}
Formally, $\hat{H}$ belongs to the universal enveloping algebra $\mathcal{E}_{\mathcal{A}}$ (UEA) of $\mathcal{A}$, which contains all polynomial functions of the Lie algebra generators
\begin{equation}
    \mathcal{E}_{\mathcal{A}} = \mathcal{A} \oplus \mathcal{A}^2 \oplus \cdots
\end{equation}
where $\mathcal{A}^N = \otimes_{n=1}^{N}\mathcal{A}$. Since $\mathcal{A}$ is compact, we can introduce the notion of the Cartan subalgebra (CSA) $\mathcal{C} \subset \mathcal{A}$, defined to be any maximal abelian Lie subalgebra of $\mathcal{A}$. A class of exactly solvable Hamiltonians are those which are linear combinations of the generators of $\mathcal{A}$
\begin{equation}
   \hat{H} = \sum_{a=1}^{d} c_a \hat{X}_a.
\end{equation}
The exact solvability follows from the maximal torus theorem, which implies that $\hat{H}$ can be transformed to an element of the Cartan subalgebra by a unitary in the corresponding Lie group. The Lie group elements take the following form
\begin{equation}
    \hat{U} = \exp\left(-i\sum_{a=1}^{d} \theta_a \hat{X}_a\right).
\end{equation}
In the context of quantum computing, exact solvability of a many-body Hamiltonian requires an additional transformation of the Cartan subalgebra generators to Pauli $\hat{z}$ operators. In practice, for the qubit-based and fermionic-based algebras considered in this work and in previous works, such transformations are Clifford unitaries.  

One can also construct exactly solvable Hamiltonians from multiple distinct Lie algebras. Let $\{\hat{X}_a^{(\lambda)} : 1 \leq a \leq d_\lambda\}$ denote a set of generators for Lie algebras $\mathcal{A}_\lambda$ with corresponding Cartan subalgebras $\mathcal{C}_\lambda$, such that generators of distinct Lie algebras commute:
\begin{equation}
    [\hat{X}_a^{(\lambda)}, \hat{X}_b^{(\sigma)}] = 0, \quad \lambda\not=\sigma.
\end{equation}
Then, the set of all operators $\{\hat{X}_a^{(\lambda)}\}$ generates a direct sum Lie algebra $\mathcal{A} = \oplus_{\lambda = 1}^{\Lambda} \mathcal{A}_\lambda$ with Cartan subalgebra $\mathcal{C} = \oplus_{\lambda=1}^{\Lambda} \mathcal{C}_\lambda$, and therefore, Hamiltonians which are linear combinations of all the generators are exactly solvable as well
\begin{align}
    \hat{H} &= \sum_{\lambda=1}^{\Lambda} \hat{H}^{(\lambda)}\nonumber\\
    \hat{H}^{(\lambda)} &= \sum_{a=1}^{d_\lambda} c_a^{(\lambda)} \hat{X}_a^{(\lambda)}.
\end{align}
The Lie group unitary that transforms $\hat{H}$ to a Cartan subalgebra element is a product of the Lie group unitaries that transform the individual $\hat{H}^{(\lambda)}$ to elements of $\mathcal{C}_\lambda$
\begin{equation}
    \hat{U} = \prod_{\lambda=1}^{\Lambda}\exp\left(-i\sum_{a=1}^{d_\lambda}\theta_a^{(\lambda)}\hat{X}_a^{(\lambda)}\right)
\end{equation}
Thus, the direct sum is an approach to construct larger Lie algebras out of smaller ones whose associated Hamiltonians can be solved efficiently. 

\subsection{\label{symmetries_subsec} Incorporation of symmetries}

Suppose there exists a set of operators which commute with all the generators $\{\hat{X}_a^{(\lambda)}\}$ of the Lie algebra $\mathcal{A}$. Then, we can define an abelian group $\mathcal{G}$ of operators with minimal generating set $G = \{\hat{C}_1,\ldots,\hat{C}_K\}$ such that
\begin{align}
    [\hat{C}_k, \hat{C}_l] &= 0\\
    [\hat{C}_k, \hat{X}_a^{(\lambda)}] &= 0,
\end{align}
for all $k, l, a, \lambda$. Then, any symmetry operator constructed as a linear combination of the symmetry group elements can be expressed as a polynomial function of the generators:
\begin{equation}
    p(\overline{C}) = \sum_{k=1}^{K} c_k \hat{C}_k + \sum_{k \geq l}^{K} c_{kl} \hat{C}_k \hat{C}_l + \cdots,
\end{equation}
where $p(\overline{C})$ is shorthand notation for $p(\hat{C}_1,\ldots,\hat{C}_K)$. As shown in Ref. \cite{patelExtensionExactlysolvableHamiltonians2023}, one can construct exactly solvable Hamiltonians by forming linear combinations of the Lie algebra generators, where the coefficients are polynomials of the symmetry group generators:
\begin{equation}
    \hat{H} = \sum_{\lambda=1}^{\Lambda} \sum_{a=1}^{d_\lambda} p_a^{(\lambda)}(\overline{C}) \hat{X}_a^{(\lambda)} + q(\overline{C}).\label{sym_aug}
\end{equation}
The exact solvability of $\hat{H}$ follows from the fact that, within each simultaneous eigenspace $E_{\vec{v}}$ of the $\hat{C}_k$, defined by states $\ket{\psi}$ such that $\hat{C}_k \ket{\psi} = v_k \ket{\psi}$, $\hat{H}$ acts effectively as an exactly solvable linear combination of the Lie algebra generators, up to a constant shift:
\begin{equation}
    \hat{H} \hat{P}_{\vec{v}} = \left[ \sum_{\lambda=1}^{\Lambda} \sum_{a=1}^{d_\lambda} p_a^{(\lambda)}(\vec{v}) \hat{X}_a^{(\lambda)} + q(\vec{v})\right] \hat{P}_{\vec{v}},
\end{equation}
where $\hat{P}_{\vec{v}}$ is a projection operator onto $E_{\vec{v}}$, and $p_a^{(\lambda)}(\vec{v})$, $q(\vec{v})$ are real numbers obtained by plugging in the associated vector of eigenvalues into the polynomials.

\subsection{\label{qubit_subsec}Constructing exactly solvable qubit Hamiltonians}

In the qubit algebra, one can construct Hamiltonians belonging to a direct sum algebra $\oplus_{\lambda=1}^{\Lambda}so(L_\lambda + 1)$ from operators $\{\hat{A}_i^{(\lambda)}\}_{i=1}^{L_\lambda}, 1 \leq \lambda \leq \Lambda$ which satisfy
\begin{align}
    \{\hat{A}_i^{(\lambda)}, \hat{A}_j^{(\lambda)}\} &= 2\delta_{ij}\\
    [\hat{A}_i^{(\lambda)}, \hat{A}_j^{(\sigma)}] &= 0, \quad \lambda\not=\sigma
\end{align}
as follows:
\begin{equation}
    \hat{H} = \sum_{\lambda=1}^{\Lambda}\left[\sum_{i=1}^{L_\lambda} c_i^{(\lambda)} \hat{A}_i^{(\lambda)} + i\sum_{i>j}^{L_\lambda} c_{ij}^{(\lambda)} \hat{A}_i^{(\lambda)}\hat{A}_j^{(\lambda)}\right]\label{so_alg_element}
\end{equation}
Setting $c_{ij}^{(\lambda)} = 0$ in the above equation produces an exactly solvable generalization of AC Hamiltonians which we will call term wise commuting AC (TWC-AC) Hamiltonians.

To use the general Lie algebra elements shown in Eq. \ref{so_alg_element} for partitioning of the electronic Hamiltonian $\hat{H}_e$, there are two complications. First, $\hat{H}_e$ is a real-symmetric Hamiltonian, implying that only real-symmetric Pauli operators can be present in the qubit representation of $\hat{H}_e$. However, any $so(L+1)$ Lie algebra obtained from Pauli operators will contain anti-symmetric generators. To demonstrate, lets consider an $so(3)$ subalgebra generated by $\{\hat{A}_i, \hat{A}_j, i \hat{A}_i \hat{A}_j\}$ for $i \not= j$. In the cases that $\hat{A}_i$ and $\hat{A}_j$ are both symmetric or both anti-symmetric, their product $i \hat{A}_i \hat{A}_j$ will be anti-symmetric, and in the case that exactly one of $\hat{A}_i$ or $\hat{A}_j$ is anti-symmetric, their product will be symmetric. Thus, any $so(3)$ subalgebra of $\oplus_{\lambda=1}^{\Lambda} so(L_\lambda + 1)$ will contain at least one anti-symmetric generator which cannot be present in $\hat{H}_e$. This limits the applicability of full Lie algebra elements shown in Eq. \ref{so_alg_element} for partitioning of the electronic Hamiltonian. 

Second, as described in the Introduction, in qubit-algebra based methods, graph partitioning algorithms are used to obtain the fragments. This requires a characterization of the exactly solvable fragment types in terms of constraints on the connectivity of their anti-compatibility graphs. However, as shown in Ref. \cite{chapmanCharacterizationSolvableSpin2020}, constraints on the connectivity of the anti-compatibility graph are not sufficient to confirm that a Hamiltonians Pauli terms belong to an $so(L+1)$ Lie algebra. To demonstrate why, suppose $\{\hat{X}_a\}_{a=1}^{d}$ are a set of Pauli operators that generate a Lie algebra, and $\{\hat{S}_a\}_{a=1}^{d}$ are a set of mutually-commuting Pauli symmetries of the $\hat{X}_a$:
\begin{equation}
    \forall a,b, \quad [\hat{S}_a, \hat{S}_b] = 0, \quad [\hat{S}_a, \hat{X}_b] = 0. \label{sym_and_so}
\end{equation}
Then, the set $\{\hat{S}_a \hat{X}_a\}_{a=1}^{d}$ has the same anti-compatibility graph as the original set $\{\hat{X}_a\}_{a=1}^{d}$, but does not necessarily generate the same Lie algebra, as
\begin{equation}
    [\hat{S}_a \hat{X}_a, \hat{S}_b \hat{X}_b] = \hat{S}_a \hat{S}_b [\hat{X}_a, \hat{X}_b],
\end{equation}
which is not in general a linear combination of the operators $\{\hat{S}_a \hat{X}_a\}_{a=1}^{d}$ since it is quadratic in the symmetries. Therefore, graph algorithms for partitioning the anti-compatibility graph of $\hat{H}_e$ will not return Hamiltonians belonging to a Lie algebra, as in Eq. \ref{so_alg_element}, but rather, Hamiltonians of the following form
\begin{equation}
    \hat{H} = \sum_{\lambda=1}^{\Lambda}\left[\sum_{i=1}^{L_\lambda} c_i^{(\lambda)} \hat{S}_i^{(\lambda)}\hat{A}_i^{(\lambda)} + i\sum_{i>j}^{L_\lambda} c_{ij}^{(\lambda)} \hat{S}_{ij}^{(\lambda)}\hat{A}_i^{(\lambda)}\hat{A}_j^{(\lambda)}\right], \label{so_sym_elements}
\end{equation}
where $\{\hat{S}_i^{(\lambda)}\} \cup \{\hat{S}_{ij}^{(\lambda)}\}$ are mutually-commuting symmetries of the Lie algebra generators $\{\hat{A}_i^{(\lambda)}\} \cup \{i\hat{A}_i^{(\lambda)} \hat{A}_j^{(\lambda)}\}$. We will call such fragments free-fermionic (FF) in the $\Lambda = 1$ case, and TWC-FF in the $\Lambda \geq 1$ case, which are generalizations of AC and TWC-AC respectively. We use the term free-fermionic since, in Ref. \cite{chapmanCharacterizationSolvableSpin2020}, it was recognized that, within each simultaneous eigenspace of the symmetry operators, $\hat{H}$ acts effectively as a one-body Majorana-fermionic Hamiltonian (which also form realizations of $so(N)$ Lie algebras). Note also that, when searching for fragments of the form shown in Eq. \ref{so_sym_elements}, it is possible to find both symmetric and anti-symmetric generators of the Lie algebra in $\hat{H}_e$, as the anti-symmetric generators can be multiplied by anti-symmetric symmetry operators to produce a symmetric Pauli term.

The FF and TWC-FF Hamiltonians are examples of linear combinations of Lie algebra generators, where the coefficients are monomials of generators of a Pauli symmetry group. As described in Section \ref{symmetries_subsec}, one can use general polynomial functions as coefficients while preserving the exact solvability. Applying this to Eq. \ref{so_alg_element} produces Hamiltonians of the following form
\begin{align}
    \hat{H} &= \sum_{\lambda=1}^{\Lambda}\left[\sum_{i=1}^{L_\lambda} p_i^{(\lambda)}(\overline{C})\hat{A}_i^{(\lambda)} + i\sum_{i>j}^{L_\lambda} p_{ij}^{(\lambda)}(\overline{C})\hat{A}_i^{(\lambda)}\hat{A}_j^{(\lambda)}\right]\nonumber\\ 
    &+ p_0(\overline{C})\label{sym_twc_ff}
\end{align}
which we call ``symmetry-augmented'' TWC-FF (Sym-TWC-FF). Analogously, we can obtain symmetry-augmented generalizations of FF, AC, and TWC-AC Hamiltonians as well by placing constraints on Eq. \ref{sym_twc_ff}. Setting $\Lambda = 1$ defines Sym-FF Hamiltonians, setting $p_{ij}^{(\lambda)}(\overline{C}) = 0$ defines Sym-TWC-AC Hamiltonians, and applying both constraints defines Sym-AC Hamiltonians. We note here that Sym-AC Hamiltonians are identical to non-contextual Hamiltonians, \cite{patelExtensionExactlysolvableHamiltonians2023} and all other symmetry augmented Hamiltonian types considered in this work are exactly-solvable generalizations of non-contextual Hamiltonians.

\subsection{\label{app_2_vqe_subsec} Using the exactly solvable Hamiltonians for measurement in VQE}

For application of the exactly solvable Hamiltonians defined in this work to measurement in VQE, three questions must be answered. First, how to obtain a partitioning of the target Hamiltonian into exactly solvable fragments of a given type. Second, how to perform any classical pre-processing necessary to express the Hamiltonian fragments in their exactly solvable form. Third, how to obtain a polynomial-sized quantum circuit for measuring the exactly solvable Hamiltonian fragments. In this section we address all three of these questions.

As described previously, to obtain a partitioning of the target Hamiltonian into solvable fragments, it is useful to have a characterization of the exactly solvable fragments in terms of constraints on the structure of their anti-compatibility graphs. In this case, the problem of obtaining the Hamiltonian decomposition can be reduced to a graph partitioning problem on the target Hamiltonians anti-compatibility graph. Therefore, as with previously considered AC, FC, and NC Hamiltonians, greedy graph-partitioning algorithms like sorted-insertion can be applied to the anti-compatibility graph of the target Hamiltonian to obtain the exactly solvable fragments. \cite{crawfordEfficientQuantumMeasurement2021} In Appendix \ref{app_a_acg_char_sec}, we provide characterizations of all exactly solvable models considered in this work in terms of the structure of the associated graphs. 

\begin{table*}[htbp]
    \centering
    {\begin{tabularx}{\textwidth}{@{\extracolsep{\fill}} c c c c c c c c c c c c c c c c c c }
        \toprule
        System  & Pauli & FC    & AC   & NC    & TWC-AC & Sym-TWC-AC & FF    & Sym-FF & TWC-FF & Sym-TWC-FF\\
        \midrule
        H$_2$   & 0.136 & 0.136 & 0    & 0     & 0.136  & 0          & 0.032 & 0       & 0.136 & 0         \\
        \midrule
        LiH     & 18.0  & 0.882 & 3.73 & 0.855 & 1.38   & 0.654      & 3.27  & 0.520   & 0.870 & 0.298     \\
        \midrule
        BeH$_2$ & 51.8  & 1.11  & 11.4 & 1.22  & 2.19   & 1.00       & 7.34  & 1.09    & 1.67  & 0.855     \\
        \midrule
        H$_2$O  & 499   & 7.59  & 126  & 9.49  & 19.3   & 4.62       & 49.4  & 4.98    & 27.0  & 5.31      \\
        \midrule
        NH$_3$  & 899   & 18.8  & 147  & 17.82 & 39.8   & 14.4      & 45.1  & 16.0    & 36.1  & 11.8      \\
        \bottomrule
        \end{tabularx}
        \caption{Comparison of variance metric $\epsilon^2 M(\epsilon)$ for using different exactly solvable qubit Hamiltonians as measurable fragments of electronic Hamiltonians. The fragment variances were calculated using the exact ground state of the corresponding electronic Hamiltonian. The variance metric for using each Pauli operator present in the Hamiltonian as its own fragment is shown in the ``Pauli'' column for comparison.}
        \label{var_table}
    }
\end{table*}

\begin{table*}[htbp]
    \centering
    {\begin{tabularx}{\textwidth}{@{\extracolsep{\fill}} c c c c c c c c c c c c c c c c c c }
        \toprule
        System  & $\hat{H}_e$& FC        & AC       & NC        & TWC-AC    & Sym-TWC-AC & FF       & Sym-FF      & TWC-FF   & Sym-TWC-FF         \\
        \midrule
        H$_2$   & 1.58(14)   & 1.38(10)  & 0.186(2) & 1.58(14)  & 1.38(10)  & 1.58(14)   & 0.24(3)  & 1.58(14)    & 1.38(10)  & 1.58(14)          \\
        \midrule
        LiH     & 13.0(630)  & 10.0(78)  & 1.15(10) & 10.2(100) & 10.0(78)  & 10.2(100)  & 1.52(29) & 10.4(126)   & 10.0(78)  & 10.4(126)         \\
        \midrule
        BeH$_2$ & 22.8(665)  & 17.6(105) & 2.38(8)  & 17.9(131) & 17.6(105) & 17.9(131)  & 2.82(22) & 18.2(161)   & 17.6(105) & 18.2(161)         \\
        \midrule
        H$_2$O  & 71.9(1085) & 55.4(105) & 12.7(11) & 57.1(131) & 55.4(105) & 57.0(137)  & 14.6(56) & 57.7(163)   & 46.5(41)  & 58.1(176)         \\
        \midrule
        NH$_3$  & 70.5(3608) & 46.1(136) & 9.35(15) & 47.4(166) & 46.1(136) & 43.0(136)  & 11.9(91) & 48.2(200)   & 39.9(88)  & 45.3(184)         \\
        \bottomrule
        \end{tabularx}
        \caption{$L_1$ norm and number of Pauli terms (shown in brackets) in the fragment with largest $L_1$ norm obtained from sorted-insertion algorithm for different exactly solvable qubit Hamiltonians. The $L_1$ norm and number of Pauli terms of the electronic Hamiltonian is presented in the $\hat{H}_e$ column for comparison.}
        \label{first_fragment_table}
    }
\end{table*}

Next, once the fragments are obtained, they are presented as a linear combination of Pauli operators as follows:
\begin{equation}
    \hat{H} = \sum_{\hat{P} \in \mathcal{S}} h_P \hat{P}.
\end{equation}
When $\hat{H}$ is one of the symmetry-augmented Hamiltonian types, measurement of $\hat{H}$ requires a classical pre-processing step which identifies the symmetries $\{\hat{C}_k\}_{k=1}^{K}$ and $\oplus_{\lambda=1}^{\Lambda} so(L_\lambda + 1)$ algebra generators from the Pauli terms $\mathcal{S}$, so that the necessary Clifford and Lie group transformations to measure $\hat{H}$ can be constructed. The procedure for performing this factorization for Sym-TWC-FF Hamiltonians is described in Appendix \ref{app_b_factorization_sec}.

Lastly, to measure an exactly solvable fragment $\hat{H}$, one must find the unitaries necessary to transform to the $\hat{H}$ eigenbasis. The procedure for measuring a symmetry-augmented Hamiltonian in which the symmetries and Lie algebra generators have already been factorized is given in Ref. \cite{patelExtensionExactlysolvableHamiltonians2023}, and here we present a summary of the main ideas when applied to Sym-TWC-FF Hamiltonians (i.e: Hamiltonians of the form shown in Eq. \ref{sym_twc_ff}). The first step is to apply a Clifford unitary $\hat{U}_c$ to transform the independent symmetries $\hat{C}_k$ of $\hat{H}$ to Pauli operators $\hat{z}_k$ on the first $K$ qubits. Then, since the Lie algebra generators commute with the symmetries, the transformed Hamiltonian $\hat{U}_c \hat{H} \hat{U}_c^\dagger$ will act only with $\hat{z}$ or $\hat{I}$ on the first $K$ qubits. Thus, these qubits can be measured, yielding eigenvalues $\vec{v} \in \{-1,1\}^{K}$ of $\hat{z}_1,\ldots,\hat{z}_K$. This has the effect of projecting the Hamiltonian to an effective Hamiltonian $\hat{H}_{\vec{v}}$ that acts on the remaining qubits. As described in Section \ref{symmetries_subsec}, this Hamiltonian belongs to a Lie algebra, and therefore, it can be transformed to an FC Hamiltonian by a unitary in the corresponding Lie group. The resulting FC Hamiltonian can be mapped to a polynomial of $\hat{z}$ operators by another Clifford unitary $\hat{V}_c$ and measured, producing eigenvalues $\vec{w} \in \{-1,1\}^{N-K}$.  The measurement results $(\vec{v}, \vec{w})$ can then be processed by a classical computer to compute the corresponding measured eigenvalue $E_H(\vec{v}, \vec{w})$. 

\section{\label{results_sec}Results}

Here, we assess the capability of the exactly solvable Hamiltonians defined in this work to serve as measurable fragments of electronic Hamiltonians for VQE. In VQE, one must compute an estimate of the expectation value $E[\psi] = \braket{\psi|\hat{H}|\psi}$ of the target Hamiltonian $\hat{H}$ via repeated quantum measurements. As $\hat{H}$ is in general not measurable directly, a partitioning
\begin{equation}
    \hat{H} = \sum_{\alpha=1}^{N_\text{frag}} \hat{H}_\alpha
\end{equation}
of $\hat{H}$ into exactly solvable fragments $\hat{H}_\alpha$ can be used, as linearity of the expectation value implies 
\begin{equation}
    E[\psi] = \sum_{\alpha=1}^{N_\text{frag}} \braket{\psi|\hat{H}_\alpha|\psi}
\end{equation}
Therefore, one can circumvent the need to measure $\hat{H}$ directly by computing the expectation value of all fragments via measurements, and summing them up. In this scenario, each fragment $\hat{H}_\alpha$ is measured separately $m_\alpha$ times, for a total number of measurements $M = \sum_\alpha m_\alpha$. It has been shown previously that the lower bound in the number of measurements necessary to achieve an accuracy of $\epsilon$ in the estimation of $E[\psi]$ is 
\begin{equation}
    M(\epsilon) = \frac{1}{\epsilon^2} \left[\sum_{\alpha=1}^{N_\text{frag}} \sqrt{\text{Var}_{\psi}(\hat{H}_\alpha)}\right]^2, \label{var_metric}
\end{equation}
where $\text{Var}_{\psi}(\hat{H}_\alpha) = \braket{\psi|\hat{H}_\alpha^2|\psi} - \braket{\psi|\hat{H}_\alpha|\psi}^2$ is the fragment variance. \cite{crawfordEfficientQuantumMeasurement2021} The lower bound can be obtained by allocating the measurements $\{m_\alpha\}$ optimally. This makes the proportionality constant $\epsilon^2 M(\epsilon)$ a fundamental metric to gauge the efficiency of different Hamiltonian decompositions when applied to quantum simulation via VQE. 

Details of the electronic Hamiltonians considered in this work are presented in Appendix \ref{app_D_sec}. We used the Bravyi-Kitaev transformation to express the electronic Hamiltonians as a linear combination of Pauli operators. For all exactly solvable Hamiltonian types, the greedy sorted-insertion algorithm was used to generate the fragments, as this method has been shown to yield lower $\epsilon^2 M(\epsilon)$ values compared with other similar graph partitioning algorithms. \cite{crawfordEfficientQuantumMeasurement2021} All relevant graph theory algorithms necessary to perform the partitioning were done using the NetworkX Python package. \cite{SciPyProceedings_11} Table \ref{var_table} shows the variance metric $\epsilon^2 M(\epsilon)$ for all different fragment types. For all systems considered, there was a reduction in measurement cost using the Sym-TWC-AC, Sym-FF, and Sym-TWC-FF criterions compared with FC, with the overall best being Sym-TWC-FF for all systems except H$_2$O, where Sym-TWC-AC was the best. That Sym-TWC-AC is sometimes better suggests that, in general, naively applying the sorted-insertion algorithm using a more general solvable Hamiltonian class will not always give a superior result, as Sym-TWC-FF is a strictly more general class than Sym-TWC-AC but does not always outperform Sym-TWC-AC. This is most pronounced when looking at the results for TWC-AC and TWC-FF fragments. Both classes performed worse than the FC fragments, even though the TWC-AC and TWC-FF criteria are more general than the FC criterion. 

Another application is to obtain exactly solvable approximations to the electronic Hamiltonian. For example, in the contextual-subspace VQE method, the ground state of a single non-contextual fragment of the electronic Hamiltonian is obtained as an initial approximation to the exact ground state. \cite{Kirby2021_contextual_subspace} For such applications, one only requires a single Hamiltonian fragment rather than a complete decomposition. For benchmarking such applications, we looked at the size of the largest fragment obtained during the decomposition, measured in terms of the $L_1$ norm of the Pauli coefficients, as well as the total number of Pauli terms in the fragment. The results are shown in Table \ref{first_fragment_table}. Here, we see that the Sym-TWC-FF fragment has a larger or equal $L_1$ norm than to all other fragnents for all molecules except NH$_3$, where FC, NC, TWC-AC, and Sym-FF have larger $L_1$ norm. A similar trend is present for the number of Pauli terms, where Sym-TWC-FF has the most Pauli terms except for NH$_3$, where Sym-FF has the most Pauli terms. This suggests that the more complex fragment types are viable for use as exactly solvable approximations to $\hat{H}_e$. We also see a similar result to that shown in Table \ref{var_table}, where TWC-AC and TWC-FF were always either matched or outperformed in the number of Pauli operators compared with FC. We attribute this to the fact that introducing anti-commuting Pauli operators into a fragment early on in a sorted-insertion iteration places more restrictive constraints on the possible Pauli operators one can further include in the same iteration-step while preserving exact solvability. To be precise: on $N$ qubits, one can obtain FC Hamiltonians with at most $2^N$ Pauli operator terms, and AC Hamiltonians with at most $2N + 1$ Pauli operator terms, \cite{bonet-monroigNearlyOptimalMeasurement2020} which is a consequence of the fact that commutativity is preserved under forming products, whereas anti-commutativity is not preserved under products.

\section{\label{conclusion_sec}Conclusions}

In this work, we applied algebraic methods to develop new decompositions of the electronic Hamiltonian for VQE. The exact solvability of the fragments is based on combining commutativity and anti-commutativity conditions for Pauli products in a way consistent with Lie algebraic structures. These algebraic structures are revealed once the Pauli symmetries are measured. The resulting effective Hamiltonian is an element of a compact Lie algebra, and can therefore be measured by applying a unitary in the corresponding Lie group. The characterization of the solvability in terms of anti-commutation relations implies that efficient graph theory algorithms can be used to obtain the Hamiltonian fragments. Using the newly considered exactly solvable Sym-TWC-AC and Sym-TWC-FF Hamiltonians produce fragments of the electronic Hamiltonian for which the ground-state expectation value can be computed with fewer measurements compared to other exactly solvable Hamiltonian types.

There are multiple avenues for future work. One question concerns developing improvements to the graph partitioning algorithms used to obtain the decomposition. We showed in this work that usage of the sorted-insertion algorithm for TWC-AC and TWC-FF yielded dramatically inferior results to FC, even though the set of FC Hamiltonians is a subset of the set of TWC-AC and TWC-FF Hamiltonians. Therefore, in principle, the optimal TWC-AC and TWC-FF decomposition should be at least as good as the optimal FC decomposition.

Another question concerns using more general solvable Hamiltonians for decompositions than the ones considered in this work. One generalization is to consider polynomial functions of distinct Lie algebra elements rather than linear combinations. For example, if $\hat{H}^{(1)},\ldots, \hat{H}^{(\Lambda)}$ is a set of FF Hamiltonians such that $\hat{H}_\text{sum} = \sum_{\lambda=1}^{\Lambda} \hat{H}^{(\lambda)}$ is TWC-FF, then the Lie group unitary $\hat{U}$ that maps $\hat{H}_\text{sum}$ to an FC Hamiltonian also maps any Hamiltonian of the form $\hat{H}_p = p(\hat{H}^{(1)},\ldots,\hat{H}^{(\Lambda)})$ to an FC Hamiltonian, where $p$ is a polynomial function. The inclusion of higher-than-linear terms poses practical challenges for application of such Hamiltonians to quantum simulation, as the solvability of such Hamiltonians is dependent of the value of the coefficients when expressed as a linear combination of Pauli operators, implying that there cannot be a characterization of the solvability based on the anti-compatibility graph of the Hamiltonian.

Lastly, there have been recent developments in constructing more general qubit Hamiltonians that can be solved via a mapping to free-fermions. \cite{fendleyFreeFermionsDisguise2019,elmanFreeFermionsDisguise2021a, chapmanUnifiedGraphTheoreticFramework2023} Such Hamiltonians also have a simple graph-theoretical characterization which can be identified in polynomial time, implying that the Hamiltonian partitioning problem can be easily solved using the sorted-insertion algorithm with these more general exactly solvable Hamiltonian types as well. For these Hamiltonians, the Majorana quadratic $so(N)$ generators are not associated to Pauli terms but rather linear combinations of many Pauli operators. Thus, construction of the diagonalizing unitary is more complicated, and would likely yield higher depth measurement circuits than those for the solvable Hamiltonians considered in this work.

\section{\label{acknowledgements_sec}Acknowledgements}

S.P. and A.F.I. are grateful to the Natural Sciences and Engineering Research Council of Canada for financial support. 

\appendix

\section{\label{app_a_acg_char_sec}Anti-compatibility graph characterization of solvable qubit Hamiltonians}

In this section, we characterize the various classes of solvable Hamiltonians considered in this work in terms of their anti-compatibility graphs, which is necessary to apply graph partitioning algorithms to obtain the exactly solvable fragments. We first start with the simplest solvable Hamiltonians defined in terms of mutually commuting or mutually anti-commuting Pauli operators. Then, we discuss the graph characterization of Hamiltonians whose Pauli operators generate an $so(N)$ Lie algebra. In the last section, we describe how incorporation of symmetries-as-coefficients can be characterized in terms of introducing additional structure to the anti-compatibility graph. For simplicity in what follows, when we refer to a Hamiltonian's graph, we mean its anti-compatibility graph. 

\subsection{\label{app_a_fc_ac_graph_subsec} Hamiltonians formed from commuting and anti-commuting Pauli operators}

\begin{figure}[h]
    \centering
    \includegraphics[width=\columnwidth]{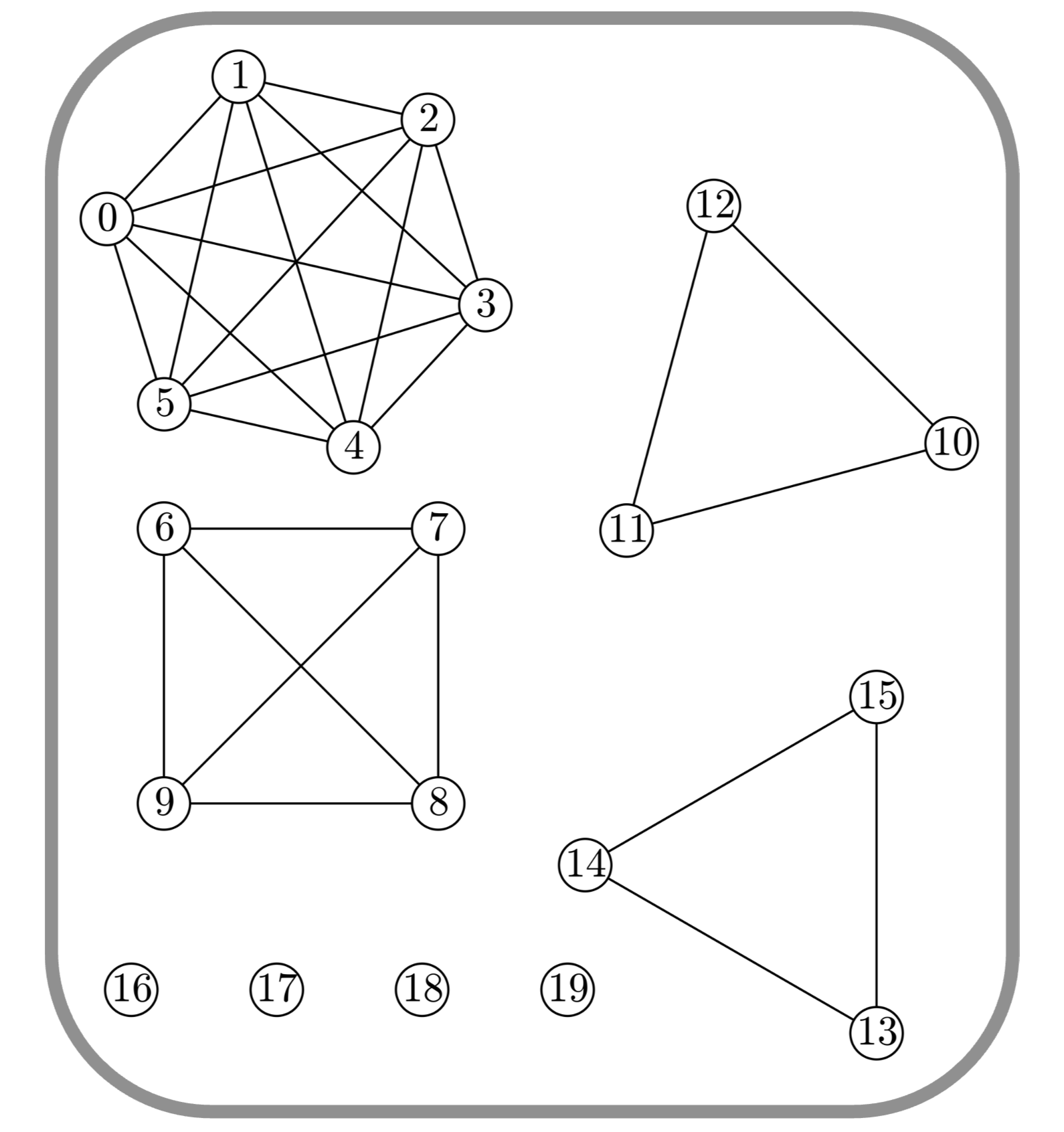}
    \caption{Example of an anti-compatibility graph of a TWC-AC Hamiltonian with $20$ Pauli terms which is a sum of 8 AC Hamiltonians. Note that the Pauli terms corresponding to the edge-less vertices 16-19 are also Pauli symmetries.}
    \label{figure1}
\end{figure}

An FC Hamiltonian formed from mutually-commuting Pauli operators will have no edges in its graph. Therefore, a Hamiltonian $\hat{H}$ is FC if and only if its graph is an empty graph. Similarly a Hamiltonian $\hat{H}$ is AC if and only if its graph is a complete graph, meaning that every pair of vertices shares an edge between them. We also defined TWC-AC Hamiltonians which are a sum of $\Lambda$ distinct AC Hamiltonians $\hat{H} = \sum_{\lambda=1}^{\Lambda} \hat{H}^{(\lambda)}$, where Pauli terms from distinct $\hat{H}^{(\lambda)}$ all mutually-commute. Then, the graph of each $\hat{H}^{(\lambda)}$ is a clique (i.e: complete subgraph) in the graph of $\hat{H}$, and all the cliques are disjoint, meaning that there are no edges shared amongst vertices in different cliques. Thus, $\hat{H}$ is TWC-AC if and only if the connected-components of its graph are all cliques. See Fig. \ref{figure1} for an example of an anti-compatibility graph of a TWC-AC Hamiltonian.

\subsection{\label{app_a_so_graph_subsec} Hamiltonians formed from $so(N)$ Lie algebra generators}

\begin{figure}[h]
    \centering
    \includegraphics[width=\columnwidth]{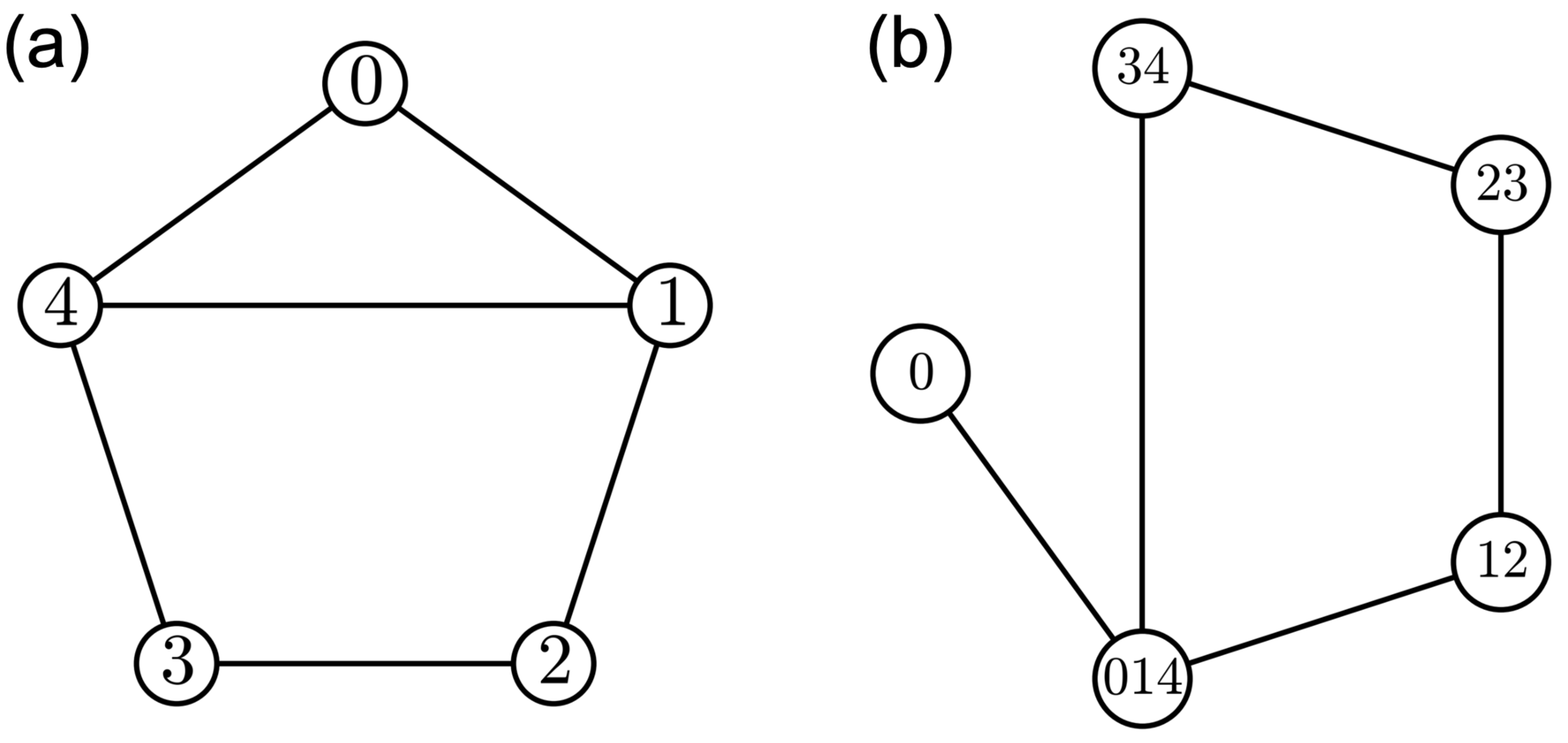}
    \caption{Example of a line graph (a) and its associated root graph (b). The vertices of the root graph are labelled by a list of vertices in the line graph which define its Krausz decomposition. For example, the vertex 014 in the root graph is associated to the clique defined by vertices 0,1, and 4 in the line graph. The vertex $0$ in the line graph is an example of a vertex that belongs to an empty clique in its Krausz decomposition.}
    \label{figure2}
\end{figure}

The graph characterization is more complex for Hamiltonians whose Pauli terms generate a full $so(N)$ Lie algebra. In Ref. \cite{chapmanCharacterizationSolvableSpin2020}, it was shown that the graph of a Hamiltonian whose Pauli terms generate an $so(N)$ Lie algebra is a line graph. Line graphs were introduced in Ref. \cite{whitneyCongruentGraphsConnectivity1932}, and the question of whether a graph $G = (V,E)$ is a line graph can be answered in $O(|E|)$ time, \cite{roussopoulosMaxAlgorithmDetermining1973, lehotOptimalAlgorithmDetect1974, degiorgiDynamicAlgorithmLine1995} implying that obtaining a partitioning of the target Hamiltonians anti-compatibility graph into line subgraphs can be done efficiently. Line graphs have many equivalent definitions, but the most relevant is the definition in terms of the so-called Krausz decomposition. \cite{krausz1943demonstration} A graph $G = (V,E)$ is a line graph if and only if its edge set $E$ can be partitioned into disjoint cliques $C_1,\ldots,C_N$, such that every vertex $v \in V$ belongs to at most two cliques.  If we allow for the possibility of empty cliques which have no edges, then we can claim that every vertex $v \in V$ belongs to exactly two cliques, which we assume has been done in what follows. Given such a clique decomposition, we can define a new graph $R = (\tilde{V}, \tilde{E})$, called the root graph of $G$, in which we associate, to every clique $C_i$ of $G$, a vertex $\tilde{v}_i$ in $\tilde{V}$, and, to every vertex $v_a$ in $V$, an edge $\tilde{e}_a$ in $\tilde{E}$. The connection of the edges $\tilde{e}_a$ to the vertices $\tilde{v}_i$ matches the membership of the vertices $v_a$ to the cliques $C_i$: for a vertex $v_a$ that belongs to cliques $C_i, C_j$, the corresponding edge $\tilde{e}_a$ connects vertices $\tilde{v}_i, \tilde{v}_j$. See Fig. \ref{figure2} for an example of a line graph and the construction of its root graph via the Krausz decomposition. For a Hamiltonian whose anti-compatibility graph is a line graph, every Pauli term is associated to a unique edge in the root graph, and the connectivity of the edges to the vertices encodes the association of the Pauli terms to $so(N)$ generators. However, as shown in Section \ref{qubit_subsec}, and described in Ref. \cite{chapmanCharacterizationSolvableSpin2020}, the converse is not true, as the anti-compatibility graph does not change when each Lie algebra generator is multiplied by a Pauli symmetry. Thus, a Hamiltonian is FF if and only if its graph is a line graph, implying that it has the form shown in Eq. \ref{so_sym_elements} with $\Lambda = 1$. Similarly, $\hat{H}$ is a TWC-FF Hamiltonian if and only if the connected-components of its graph are line graphs. 

\subsection{\label{app_a_sym_graph_subsec} Incorporation of Symmetries}

\begin{figure*}
    \centering
    \includegraphics[width=\textwidth]{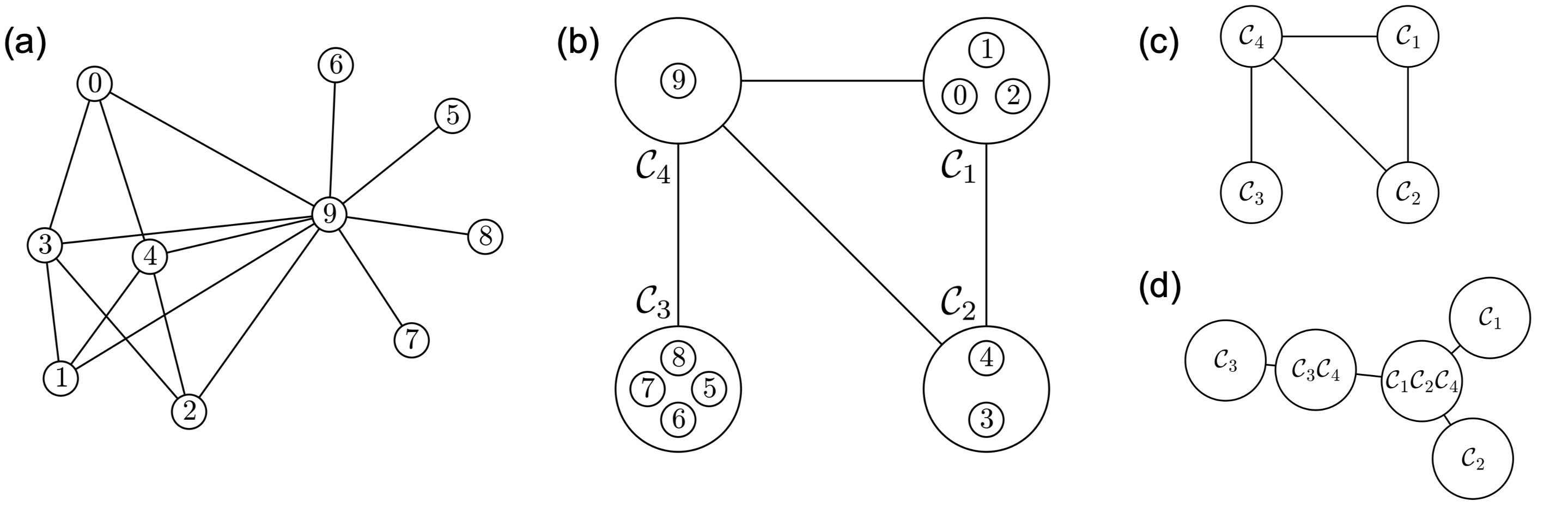}
    \caption{Example derivation of a twin-free quotient graph. In (a), the initial graph is depicted with the vertices enumerated. The first step is to partition the vertices into equivalence classes $\mathcal{C}_i$ of mutually-twin vertices, as shown in (b). In (b), the equivalence classes are grouped together, revealing the structure of the quotient graph shown in (c). In (b) and (c), the edges connecting groups $\mathcal{C}_i$ and $\mathcal{C}_j$ imply that all vertices in $\mathcal{C}_i$ share an edge with all vertices in $\mathcal{C}_j$. In (d), the root graph of the quotient graph is depicted, implying that the quotient graph is a line graph.}
    \label{figure3}
\end{figure*}

Lastly, we can provide a graph-theoretical characterization of how substitution of symmetry-polynomials into the coefficients of a qubit Hamiltonian modifies its graph:
\begin{equation}
    \hat{H} = \sum_a c_a \hat{X}_a \mapsto \hat{H}_\text{sym} = \sum_a p_a(\overline{C}) \hat{X}_a + p_0(\overline{C}).\label{sym_gen}
\end{equation}
To do this, we define the notion of twin Pauli operators. Given a set $\mathcal{S}$ of Pauli operators with corresponding graph $G$, we say that $\hat{P}$ and $\hat{Q}$ in $\mathcal{S}$ are twin Pauli operators if for all $\hat{R} \in \mathcal{S}$, $[\hat{P},\hat{R}] = 0$ if and only if $[\hat{Q}, \hat{R}] = 0$. We call such Pauli operators twins since they correspond to the graph-theoretical notion of twin vertices in the graph $G$, defined to be any pair of vertices in $G$ which share an edge with the exact same set of vertices in $G$. The connection between twin vertices in the graph of a Hamiltonian and the existence of Pauli symmetries was originally described in Ref. \cite{chapmanCharacterizationSolvableSpin2020}. Twin Pauli operators satisfy a few properties relevant to the following discussion. First, any two twin Pauli operators must commute. Moreover, if $\hat{P}$ and $\hat{Q}$ are twins, then their product $\hat{P}\hat{Q}$ is a Pauli operator that commutes with all Pauli operators in $\mathcal{S}$, since, if $\hat{R} \in \mathcal{S}$, then $\hat{P}$ and $\hat{Q}$ either both commute with $\hat{R}$, or $\hat{P}$ and $\hat{Q}$ both anti-commute with $\hat{R}$, implying their product must commute with $\hat{R}$. Lastly, the property of being twin Pauli operators is an equivalence relation on $\mathcal{S}$, implying that $\mathcal{S}$ partitions into disjoint equivalence classes $\mathcal{C}_i$ of mutually twin Pauli operators. This partitioning allows us to define a new graph $G'$, which we call the twin-free quotient graph of $G$, in which the vertices correspond to the equivalence classes of $\mathcal{S}$, and such that two vertices $i$ and $j$ in $G'$ share an edge if and only if the Pauli operators in $\mathcal{C}_i$ anti-commute with the Pauli operators in $\mathcal{C}_j$. $G'$ is isomorphic to the anti-compatibility graph of a set of Pauli operators $\{\hat{P}_i\}$ obtained by selecting one element of each equivalence class. In Fig. \ref{figure3}, we present an example of the partitioning of the vertices of a graph into subsets of mutually-twin vertices, and the derivation of the resulting twin-free quotient graph. 

The relevance of the above to characterizing the graph of exactly solvable Hamiltonians $\hat{H}_\text{sym}$ of the form shown in Eq. \ref{sym_gen} is as follows. If $\mathcal{S}$ denotes the set of Pauli operators in $\hat{H}_\text{sym}$, then each term $p_a(\overline{C}) \hat{X}_a$ and $p_0(\overline{C})$ in $\hat{H}_\text{sym}$ are linear combinations of twin Pauli operators in $\mathcal{S}$. Moreover, if one removes the term $p_0(\overline{C})$, or, equivalently, the edge-less vertices in the graph of $\hat{H}_\text{sym}$, then the twin-free quotient graph of the resulting Hamiltonian $\sum_a p_a(\overline{C}) \hat{X}_a$ is isomorphic to the anti-compatibility graph of the Hamiltonian $\hat{H} = \sum_a c_a \hat{X}_a$. Thus, this allows us to characterize the structure of an exactly solvable Hamiltonian $\hat{H}_\text{sym} = \sum_a p_a(\overline{C}) \hat{X}_a + p_0(\overline{C})$ in terms of the structure of a simpler exactly solvable Hamiltonian $\sum_a c_a \hat{X}_a$. To be precise, a Hamiltonian $\hat{H}$ is Sym-TWC-FF if and only if: (1) after removal of the fully-commuting operators in $\hat{H}$, and (2) computing the twin-free quotient graph of the resulting Hamiltonian, the result is a graph of a TWC-FF Hamiltonian. One can analogously obtain graph characterizations of Sym-AC (i.e: NC), Sym-FF, and Sym-TWC-AC Hamiltonians. In Fig. \ref{figure3}a, a graph is depicted whose twin-free quotient graph, shown in Fig. \ref{figure3}c, is a line graph, implying that any Hamiltonian whose anti-compatibility graph is isomorphic to the graph shown in Fig. \ref{figure3}a is an exactly solvable Sym-FF Hamiltonian.

\section{\label{app_b_factorization_sec}Factorization of Sym-TWC-FF Hamiltonians}

Let $\hat{H}$ denote a Sym-TWC-FF Hamiltonian obtained as a fragment of a target Hamiltonian, which we express in the following form
\begin{equation}
    \hat{H} = \sum_{\hat{P} \in \mathcal{S}} h_P \hat{P}.
\end{equation}
Measurement of $\hat{H}$ on a quantum computer requires performing a classical pre-processing step in which the Pauli terms in $\mathcal{S}$ are factorized so that $\hat{H}$ is expressed in the following form
\begin{equation}
    \hat{H} = \sum_{\lambda=1}^{\Lambda}\sum_{a=1}^{d_\lambda} p_a^{(\lambda)}(\overline{C}) \hat{X}_a^{(\lambda)} + p_0(\overline{C}),\label{sym_twc_ff_factorized}
\end{equation}
where $\{\hat{X}_a^{(\lambda)}\}$ generate a $\oplus_{\lambda=1}^{\Lambda} so(N_\lambda)$ Lie algebra, and $[\hat{C}_k, \hat{C}_l] = [\hat{C}_k, \hat{X}_a^{(\lambda)}] = 0$ for all $k, l, a, \lambda$. The procedure presented here for factorizing a Sym-TWC-FF Hamiltonian is a generalization of the procedure originally developed to factorize NC Hamiltonians introduced in Ref. \cite{kirbyClassicalSimulationNoncontextual2020}, which exploits the anti-compatibility graph characterizations described in Appendix \ref{app_a_acg_char_sec}. To proceed, we partition the set $\mathcal{S}$ into equivalence classes of mutually twin Pauli operators. The first class is the set $\mathcal{Z} \subset \mathcal{S}$ of Pauli operators that commute with all Pauli operators in $\mathcal{S}$. To obtain the remaining classes, we first partition $\mathcal{S} \setminus \mathcal{Z} = \bigcup_{\lambda=1}^{\Lambda} \mathcal{T}^{(\lambda)}$ into disjoint subsets $\mathcal{T}^{(\lambda)}$ whose graphs correspond to the connected-components in the graph of $\mathcal{S} \setminus \mathcal{Z}$. Then, we partition each $\mathcal{T}^{(\lambda)} = \bigcup_{a=1}^{d_\lambda} \mathcal{C}_a^{(\lambda)}$ into the equivalence classes of mutually twin Pauli operators. Enumerating the classes:
\begin{align}
    \mathcal{Z} &= \{\hat{Z}_\alpha : 1 \leq \alpha \leq |\mathcal{Z}|\}\\
    \mathcal{C}_a^{(\lambda)} &= \{\hat{A}_{ab}^{(\lambda)} : 1 \leq b \leq |\mathcal{C}_a^{(\lambda)}|\},
\end{align}
we can write the Sym-TWC-FF Hamiltonian in the following form
\begin{equation}
    \hat{H} = \sum_{\alpha=1}^{|\mathcal{Z}|} h_\alpha \hat{Z}_\alpha + \sum_{\lambda=1}^{\Lambda} \sum_{a=1}^{d_\lambda} \sum_{b=1}^{|\mathcal{C}_a^{(\lambda)}|} h_{ab}^{(\lambda)} \hat{A}_{ab}^{(\lambda)}.
\end{equation}
The first step to factorizing the Pauli terms in $\hat{H}$ is to factor a single element of each class $\mathcal{C}_a^{(\lambda)}$. Define $\hat{A}_a^{(\lambda)} = \hat{A}_{a1}^{(\lambda)}$ and $\hat{C}_{ab}^{(\lambda)} = \hat{A}_{ab}^{(\lambda)} \hat{A}_{a}^{(\lambda)}$, so that $\hat{H}$ can be written as
\begin{equation}
    \hat{H} = \sum_{\alpha=1}^{|\mathcal{Z}|} h_\alpha \hat{Z}_\alpha + \sum_{\lambda=1}^{\Lambda} \sum_{a=1}^{d_\lambda} \left[\sum_{b=1}^{|\mathcal{C}_a^{(\lambda)}|} h_{ab}^{(\lambda)} \hat{C}_{ab}^{(\lambda)}\right]\hat{A}_{a}^{(\lambda)}.
\end{equation}
To proceed further, we note two relevant facts. First, the Pauli operators $\{\hat{C}_{ab}^{(\lambda)}\}$ commute with all Pauli operators in $\hat{H}$, which follows from the fact that each $\hat{C}_{ab}^{(\lambda)}$ is a product of two twin Pauli operators. Second, since $\{\hat{A}_a^{(\lambda)}\}$ is obtained by sampling a single Pauli operator from each equivalence class $\mathcal{C}_a^{(\lambda)}$, the anti-compatibility graph of $\{\hat{A}_a^{(\lambda)}\}$ is the twin-free quotient graph of the anti-compatibility graph of $\mathcal{S} \setminus \mathcal{Z}$, which, since $\hat{H}$ is a Sym-TWC-FF Hamiltonian, is a graph whose connected-components are line graphs. Thus, the $\{\hat{A}_a^{(\lambda)}\}$ generate a direct sum of $so(N_\lambda)$ Lie algebras up to a set of mutually-commuting Pauli symmetries which must be factored out, producing the following factorization:
\begin{equation}
    \hat{A}_a^{(\lambda)} = \hat{S}_a^{(\lambda)} \hat{X}_a^{(\lambda)},
\end{equation} 
where $[\hat{S}_a^{(\lambda)}, \hat{S}_b^{(\sigma)}] = 0$ and $[\hat{S}_a^{(\lambda)}, \hat{A}_b^{(\sigma)}] = 0$ for all $a,b,\lambda,\sigma$. In general, Pauli symmetries can be obtained efficiently via standard linear-algebraic methods developed for qubit tapering and for solving FC Hamiltonians. \cite{bravyiTaperingQubitsSimulate2017, yenMeasuringAllCompatible2020} However, for these Hamiltonians, it was shown in Ref. \cite{chapmanCharacterizationSolvableSpin2020} that, for Pauli terms whose anti-compatibility graphs are line graphs, graph theoretical techniques can also be used to obtain not only the symmetry operators $\hat{S}_a^{(\lambda)}$, but also an expression of the symmetry operators as products of the Pauli terms which generate the line graph:
\begin{equation}
    \hat{S}_a^{(\lambda)} = \prod_{k=1}^{K_{a,\lambda}} \hat{A}_{a_k}^{(\lambda)},
\end{equation}
from which it also follows that $[\hat{S}_a^{(\lambda)}, \hat{C}_{ab}^{(\lambda)}] = 0$ and $[\hat{S}_a^{(\lambda)}, \hat{Z}_\alpha] = 0$. This factorization produces:
\begin{equation}
    \hat{H} = \sum_{\alpha=1}^{|\mathcal{Z}|}h_\alpha \hat{Z}_\alpha + \sum_{\lambda=1}^{\Lambda} \sum_{a=1}^{d_\lambda} \left[\sum_{b=1}^{|\mathcal{C}_a^{(\lambda)}|} h_{ab}^{(\lambda)} \hat{C}_{ab}^{(\lambda)} \hat{S}_a^{(\lambda)}\right]\hat{X}_a^{(\lambda)},\label{sym_twc_ff_almostfactorized}
\end{equation}
where $\{\hat{X}_a^{(\lambda)}\}$ generate a $\oplus_{\lambda=1}^{N_\lambda} so(N_\lambda)$ Lie algebra. The set of operators $\{\hat{Z}_\alpha\} \cup \{\hat{C}_{ab}^{(\lambda)}\} \cup \{\hat{S}_a^{(\lambda)}\}$ are a mutually commuting set of Pauli operators that also commute with the $\{\hat{X}_a^{(\lambda)}\}$, and can therefore all be expressed as products, up to a phase, of some minimal generating set $G = \{\hat{C}_1,\ldots,\hat{C}_K\}$ of mutually-commuting independent Pauli operators. Then, the linear combinations of the Pauli operators $\{\hat{Z}_\alpha\} \cup \{\hat{C}_{ab}^{(\lambda)}\} \cup \{\hat{S}_a^{(\lambda)}\}$ present in $\hat{H}$ can be expressed as polynomials over $G$:
\begin{align}
    \sum_{\alpha=1}^{|\mathcal{Z}|}h_\alpha \hat{Z}_\alpha &= p_0(\overline{C})\\
    \sum_{b=1}^{|\mathcal{C}_a^{(\lambda)}|} h_{ab}^{(\lambda)} \hat{C}_{ab}^{(\lambda)} \hat{S}_a^{(\lambda)} &= p_a^{(\lambda)}(\overline{C}).
\end{align}
Substitution of these polynomials into Eq. \ref{sym_twc_ff_almostfactorized} produces an expression of $\hat{H}$ in the form shown in Eq. \ref{sym_twc_ff_factorized}.

\section{\label{app_D_sec}Details of electronic Hamiltonians}

All electronic Hamiltonians used in this work were generated in the STO-3G basis and using the Bravyi-Kitaev transformation, as implemented in the Openfermion \cite{mccleanOpenFermionElectronicStructure2020} package using PySCF. \cite{sunPySCFPythonbasedSimulations2018, sunRecentDevelopmentsPySCF2020} The nuclear geometries used were:
\begin{itemize}
    \item H$_2$: R$(\rm H-H) = 1 \AA$ 
    \item LiH:  R$(\rm Li-H) = 1 \AA$ 
    \item BeH$_2$:  R$(\rm Be-H) = 1 \AA$ with colinear atomic arrangement
    \item H$_2$O: R$(\rm O-H) = 1 \AA$ with $\angle \rm HOH = 107.6^\circ$ 
    \item NH$_3$: R$(\rm N-H) = 1 \AA$ with $\angle \rm HNH = 107^\circ$ 
\end{itemize}

\bibliography{library}
\end{document}